\documentclass[twocolumn,prl,preprintnumbers,amsmath,superscriptaddress,amssymb,showpacs]{revtex4}
\usepackage[latin9]{inputenc}
\usepackage{bm}
\usepackage{amssymb}
\usepackage{graphicx}
\usepackage{esint}
\PassOptionsToPackage{normalem}{ulem}
\usepackage{ulem}

\makeatletter
{%
\usepackage{dcolumn}% Align table columns on decimal point
\usepackage{bm}% bold math
\usepackage{subfigure}%\nofiles

\makeatother

\begin{document}

\title{Gap Generation in Topological Insulator Surface States by non-Ferromagnetic Magnets}

\author{László Oroszlány}

\affiliation{Department of Physics of Complex Systems, Eotvos University, H-1117
Budapest, Pázmány Péter sétány 1/A, Hungary}

\author{Alberto Cortijo}

\affiliation{Instituto de Ciencia de Materiales de Madrid, CSIC, Cantoblanco,
28049 Madrid, Spain.}
\begin{abstract}
It is shown that, contrary to the naive expectation, single particle
spectral gaps can be opened on the surface states of three dimensional
topological insulators by using commensurate out- and in-plane antiferromagnetic
or ferrimagnetic insulating thin films. 
\end{abstract}
\maketitle
\emph{Introduction.} One of the most remarkable properties of a three
dimensional topological insulator is the presence of a topologically
quantized magnetoelectric term (TMET) in its electromagnetic response.
This term has far reaching consequences since it constitutes a condensed-matter
realization of axion electrodynamics\cite{W87,K09}. Experimental
signatures of the TMET include the quantized Kerr angle and Faraday
rotation\cite{TM10,MQD10,JSS10}, Casimir repulsion\cite{GC11}, inverse
spin galvanic effect\cite{GF10}, monopole images\cite{QLZ09}, surface
half integer Hall effect\cite{QHZ08}, topological viscoelastic response\cite{HLF11}
just to name a few.

The key point for the observability of this topologically quantized
response is the breakdown of the time reversal symmetry in the surface
of the otherwise time reversal invariant TI\cite{QHZ08}. In terms
of the electric and magnetic fields, the TMET in the electromagnetic
action takes the form

\begin{equation}
\mathcal{S}_{\theta}=\frac{\alpha}{4\pi^{2}}\int d^{3}\mathbf{r}dt\theta\mathbf{E}\cdot\mathbf{B},
\end{equation}
where $\alpha$ is the fine structure constant and $\theta$ is the
so called axion parameter which takes the value of $0$ or $(2n+1)\pi$
with $n\in\mathbb{N}$ in trivial and topological insulators, respectively\cite{QHZ08}.
Alternatively to the above description in terms of the electromagnetic
fields, one can understand the TMET as a Chern Simons (CS) term induced
in the electromagnetic response of the insulator by the gapped surface
states of a TI that are described by usual massive Dirac Hamiltonian:
\begin{equation}
H_{D}=v\left(\bm{\sigma}\times\mathbf{k}\right)\cdot\hat{\mathbf{z}}+m\sigma_{z},\label{masslessDirac}
\end{equation}
where $v$ is the Fermi velocity, and $m$ is the induced mass of
the Dirac states. In this case, the value $\theta=\pi$ corresponds
to the value $\sigma=\frac{1}{2}$sign$(m)$ for the Hall coefficient
in the corresponding CS term. In short, breaking time reversal symmetry
opens a gap in the TI surface states, thus making the TMET observable.

\begin{figure}
\includegraphics[scale=0.18]{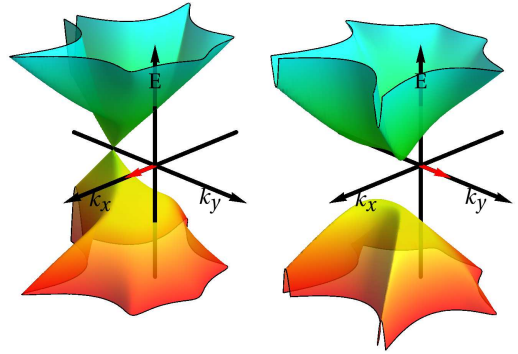} \caption{(Color online). Low energy spectrum of the surface states of a TI
in homogeneous in-plane magnetization. Red arrows show the direction
of the magnetization field.}

\label{nonmagneticbandstructure} 
\end{figure}

Within the effective low energy approximation described by (\ref{masslessDirac})
there are several proposals in the literature for opening a gap in
the helical metal by means of weak magnetic fields through a Zeeman
term $H_{Z}=g\mu_{B}\bm{\sigma}\cdot\mathbf{B}$\cite{QHX10}, or
through exchange coupling to ferromagnetic thin films $H_{exc}=J\mathbf{M}\cdot\bm{\sigma}$\cite{QHZ08},
and magnetic impurities $H_{imp}=J\sum_{j}\mathbf{S}_{j}\cdot\bm{\sigma}\delta(\mathbf{r}-\mathbf{R}_{j})$\cite{LLX09,CCA10,WXX10}.
The exchange coupling between magnetic thin films and TI is the most
appealing from the theoretical point of view because it not only gives
a simple mechanism to develop the theory of the TMET but it allow
us to look for unexpected effects that can alter the thin film magnetization
dynamics\cite{GF10,NN10}. However, this proposal is experimentally
challenging, and also it poses some questions. First of all, it is
not so easy to find insulating ferromagnetic materials. Some candidate
materials like GdN and EuO have been theoretically suggested\cite{GF10,TM10}
but to the best of our knowledge so far there is no experimental evidence
supporting this claim. Also, even if ferromagnetic insulating thin
films were available, it is not guaranteed that the thin film magnetization
would point in the out-of-plane direction\cite{GE01}. There is the
problem of possible mismatch between the TI surface lattice structure
and the thin film lattice structure and even the issue of the two
lattices not being commensurate. These problems are in the heart of
the experimental difficulties for implementing this mechanism.

\begin{figure*}
\includegraphics[scale=1.1]{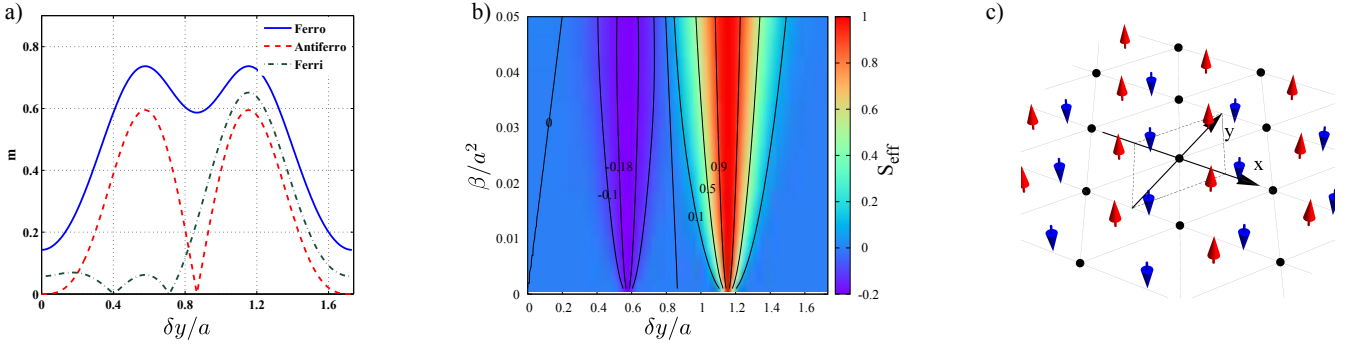} \caption{a): Exchange induced gap $m$ (in units of $B_{11}$) on the TI surface
states vs. the lattice mismatch $\delta\mathbf{y}$. Blue (solid),
red (dashed), and green (dashed-dotted) lines corresponds to ferromagnetic,
antiferromagnetic, and ferrimagnetic lattices, respectively For the
ferrimagnetic configuration $\mathbf{S}_{1}=-5\mathbf{S}_{2}$ and
$\beta=0.02a^{2}$. b): The effective Zeeman field on the surface
as the function of the relative displacement $\delta\mathbf{y}$ between
the magnetic and the TI lattices and the localization parameter $\beta$.
c): Real space configurations for the TI surface. Black dots represent
lattice Se positions and the arrows correspond to a hexagonal antiferromagnetic
lattice.}

\label{simplegaps} 
\end{figure*}

Directly using eq.(\ref{masslessDirac}) implicitly forces us to consider
a continuum medium approach for the magnetization \cite{FE12}.The
question is then how to construct this effective description of the
exchange coupling between the surface electronic spin and the magnetization
starting from a microscopic model. For ideal insulating ferromagnets
the most naive way would be to couple the electron spin with the averaged
magnetization in the magnetic unit cell. However when more realistic
magnetic insulators are considered we immediately run into difficulties.
For instance this approach automatically rules out the possibility
of considering antiferromagnetic insulators as magnetic material candidates.
Also it is not at all clear what the correct form for a continuum
description of the magnetization of a ferrimagnetic insulator is.
Motivated by these experimental and theoretical issues, we address
in this work the problem of coupling a magnetically active thin film
with the surface electronic states of a TI employing an tight binding
approach.

\emph{The model.} In order to study qualitatively the ways in which
the gap can be microscopically induced in the surface spectrum we
will employ a tight binding model valid for the topological insulators
of the form Bi$_{2}$X$_{3}$ which include the prototypical examples
of TI's Bi$_{2}$Se$_{3}$ and Bi$_{2}$Te$_{3}$. We will then follow
references \cite{LQZ10,MYK11} and will consider a Bi$_{2}$Se$_{3}$
sample made of $N$ quintuple layers (QL) grown in the $(111)$ direction
and terminated in Se planes. The surface will thus have a triangular
lattice structure. We are interested in the bandstructure around the
Fermi level so the tight-binding basis set will be made of linear
combinations of the atomic orbitals $\left(\vert p_{Bi}^{+},\uparrow\rangle,\vert p_{Se}^{-},\uparrow\rangle,\vert p_{Bi}^{+},\downarrow\rangle,\vert p_{Se}^{-},\downarrow\rangle\right)$.
The superscript reflect the parity of the state and the second index
is the spin polarization. The tight binding Hamiltonian in real space
can be written in this basis in the following way\cite{LQZ10,MYK11}:

\begin{eqnarray}
H & = & \sum_{\mathbf{n}}C_{\mathbf{n}}^{\dagger}\hat{\epsilon}C_{\mathbf{n}}+\sum_{\mathbf{n},\mathbf{a}_{i}/\mathbf{b}_{i}}C_{\mathbf{n}}^{\dagger}\hat{t}_{\mathbf{a}_{i}/\mathbf{b}_{i}}C_{\mathbf{n}+\mathbf{a}_{i}/\mathbf{b}_{i}}+H.c.\label{TBhamiltonian}
\end{eqnarray}
Here the lattice vectors $\mathbf{a}_{i}$ and $\mathbf{b}_{i}$ connect
unit cell positions within the same QL and of different QL, respectively
and $\mathbf{n}$ labels the lattice positions as defined in \cite{LQZ10,MYK11}.
We use $a=\vert\mathbf{a}_{i}\vert$ as the lateral spatial length
scale. The on-site energy $\hat{\epsilon}$ and hopping terms $\hat{t}_{\mathbf{a}_{i}/\mathbf{b}_{i}}$
are $4\times4$ matrices which can be written as linear combination
of $\Gamma_{i}$ matrices which are matrix products of spin $\bm{\sigma}$
and parity $\bm{\tau}$ Pauli matrices:

\begin{eqnarray}
 & \hat{\epsilon}=\epsilon_{0}\Gamma_{0}+m\Gamma_{5},\nonumber \\
 & \hat{t}_{\mathbf{a}_{1}}=A_{0}\Gamma_{0}-i(A_{12}\Gamma_{3}-A_{14}\Gamma_{2})+A_{11}\Gamma_{5},\nonumber \\
 & \hat{t}_{\mathbf{b}_{1}}=B_{0}\Gamma_{0}+i(B_{12}\Gamma_{4}-B_{14}\Gamma_{1})+B_{11}\Gamma_{5},\label{parameters}\\
 & \Gamma_{1,2}=\tau_{1}\otimes\sigma_{1,2},\Gamma_{3}=\tau_{1}\otimes\sigma_{3},\nonumber \\
 & \Gamma_{4,5}=\tau_{2,3}\otimes\sigma_{0},\Gamma_{0}=\tau_{0}\otimes\sigma_{0}.\nonumber 
\end{eqnarray}
The remaining hopping matrices $\hat{t}_{\mathbf{a}_{2,3}/\mathbf{b}_{2,3}}$
can be obtained from (\ref{parameters}) by applying the rotation
operation $R_{3}=\exp(i\frac{\pi}{3}\sigma_{3}\otimes\tau_{0})$.
The Hamiltonian (\ref{TBhamiltonian}) is thus made of intra-QL hopping
terms and on-site energies and hopping terms coupling different QL.
In all calculations presented here we use $B_{11}=1$, $A_{14}=1.4$,
$A_{12}=B_{12}=3$, $A_{11}=2$, $m=-10$, $B_{14}=A_{0}=B_{0}=0$,
for modelling a bulk TI\cite{MYK11}.Next we add a exchange term coupling
to the Se atomic orbitals in (\ref{TBhamiltonian}) of the first QL
of the form

\begin{equation}
H_{exc}=J\sum_{\mathbf{n}}\mathbf{S}(\mathbf{R}_{\mathbf{n}})C_{\mathbf{n}}^{\dagger}\bm{\Sigma}C_{\mathbf{n}}.
\end{equation}

where The matrices $\bm{\Sigma}$ are of the form $\bm{\Sigma}=\frac{1}{2}\left(\tau_{0}-\tau_{3}\right)\otimes\bm{\sigma}$
and now $\mathbf{R}_{\mathbf{n}}$ represent the lattice positions
of the Se atoms in the outer part of the first QL. The important observation
here is that the magnetic and surface lattices \emph{do not need to
be the same for generic magnetic layers} so the magnetic moment of
the magnetic layer $\mathbf{S}(\mathbf{R}_{\mathbf{n}})$ at $\mathbf{R}_{\mathbf{n}}$
will not be the magnetic moment of each magnetic position. Usually
the magnetic moments represent the magnetic moment associated to a
bounded atomic orbital with a short spatial extension so not all the
magnetic moments will couple in the same manner to the electronic
spins on the surface and the coupling will be stronger for nearer
atoms. The two previous observations lead us to define $\mathbf{S}(\mathbf{R}_{\mathbf{n}})$
as 
\begin{equation}
\mathbf{S}(\mathbf{R}_{\mathbf{n}})\equiv\sum_{i}\mathbf{S}(\mathbf{\hat{R}}_{i})\Phi\left(\mathbf{R}_{\mathbf{n}}-\mathbf{\hat{R}}_{i}\right),\label{effmoment}
\end{equation}
where now the sum is performed over the magnetic lattice positions.
The function $\Phi$ encodes the information about the short range
character of the localized magnetic orbitals. In our calculations
we have chosen a Gaussian profile, $\Phi(\mathbf{r})=e^{-\mathbf{r}^{2}/\beta}$
parametrized by the parameter $\beta$ which has the meaning of the
(squared) mean size of the spatial profile of the magnetic orbital.
We have checked that any other choice for $\Phi$ does not modify
the qualitative results presented in this work.It is important to
note that for a given Se position, nearby moments will contribute
to $\mathbf{S}(\mathbf{R}_{\mathbf{n}})$ but not equally if there
is a mismatch between the two sublattices. This key observation is
interesting because it opens the possibility of considering not just
ferromagtetic, but also other types of magnetic ordering as a candidate
for inducing gaps in the TI surface states by the exchange coupling
mechanism. 

\emph{Results.} In order to show the ideas explained above at work,
let us consider first the case where the positions of the magnetic
lattice lie in the middle of the triangles formed by the surface lattice
as it is shown in Fig.(\ref{simplegaps}c) and calculate the spectrum
with eqs (\ref{TBhamiltonian}-\ref{effmoment}). We will control
the mismatch between the lattices by displacing a magnetic bipartite
lattice a distance $\delta\mathbf{y}$ with respect the center of
the triangle in the $OY$ direction, and we will consider the three
cases of ferromagnetic ($\mathbf{S}_{1}=\mathbf{S}_{2}$), antiferromagnetic
($\mathbf{S}_{1}=-\mathbf{S}_{2}$) and ferrimagnetic ($\mathbf{S}_{1}=-5\mathbf{S}_{2}$)
out of plane configurations. In Fig.(\ref{simplegaps}a) we show the
value of the gap defined as $m=\vert\min[E_{c}(\mathbf{k})]-\max[E_{v}(\mathbf{k})]\vert/2$.
For the ferromagnetic case, the system always develops a non zero
gap, as expected, irrespective of the relative position of the two
lattice sites. The modulation in the value of the gap is understood
in terms of the different contribution of the magnetic moments to
$\mathbf{S}_{eff}(\mathbf{R}_{j})$. Much more interesting are the
cases of ferrimagnetic and antiferromagnetic lattice structures. The
first important observation is that in both cases a gap is opened
when varying the relative position of the lattices, showing that in
principle one can open gaps in the TI surface states by the interaction
with ferri- and antiferromagnetic layers. In principle, nothing guarantees
that the magnetic lattice sites must lie on the exact center of the
triangles formed by the surface positions, but the gap might be still
open. Moreover, if during the fabrication process it were possible
to control the lattice mismatch, the gap could be tuned. Another important
observation is that although $m$ is a positive definite quantity
by construction, the value of the effective Zeeman coupling is not.
Indeed it will change its sign, as it is shown in fig.(\ref{simplegaps}b),
where the effective Zeeman term is plotted as a function of the lattice
displacement $\delta\mathbf{y}$ and the value of $\beta$. This result
means that for the case of ferrimagnet different relative displacements
might lead to different values of the coupling, which is the signature
of a topological phase transition, controlled by $\delta\mathbf{y}$.
This is also true for generic antiferromagnetic configurations. As
can be readily seen in fig.(\ref{simplegaps}b) there is always a
change of sing of $\mathbf{S}_{eff}$ irrespective of how tight are
the magnetic atomic orbitals to their lattice sites.

So far, we have considered magnetic configurations in thin layers
with the magnetization being out-of-plane. It is well known that when
thin film geometries are considered for ferromagnets it is more energetically
favorable for the system to have the magnetization in plane\cite{BM88,GE01}.
From the form of (\ref{masslessDirac}) an in-plane homogeneous magnetization
would not induce any gap since an in plane magnetic moment would just
shift the position of the Dirac point. Actually this is not the case
and a gap can be induced when lattice effects are considered in addition
to (\ref{masslessDirac}) as it was shown by Fu\cite{F09}. We can
add to (\ref{masslessDirac}) the two next to leading terms in the
expansion in momenta\cite{F09}: 
\begin{equation}
H_{w}=\frac{\mathbf{k}^{2}}{2m_{0}}\sigma_{0}+\alpha\mathbf{k}^{2}\left(\bm{\sigma}\times\mathbf{k}\right)\cdot\hat{\mathbf{z}}+\lambda\left(k_{x}^{3}-3k_{x}k_{y}^{2}\right)\sigma_{z},\label{hexwarping}
\end{equation}
where $m_{0}$, $\alpha$, and $\lambda$ come from the comparison
between the band structure calculated from the tight binding and ARPES
measurements\cite{C09}. When the Hamiltonian $H_{D}+H_{w}$ is considered
together with $H_{exc}=J_{\parallel}\bm{\sigma}_{\parallel}\mathbf{m}_{\parallel}$,
it is apparent that a gap of value $\tilde{m}=\lambda\frac{J_{\parallel}^{3}}{v^{3}}\left(m_{y}^{3}-3m_{y}m_{x}^{2}\right)$
appears at $\mathbf{k}_{g}=\frac{J_{\parallel}}{v}\mathbf{m}_{\parallel}\times\hat{\mathbf{z}}$
as it is shown in fig.(\ref{nonmagneticbandstructure}). Apart from
the mass generation due to the hexagonal terms there is a self-doping
effect defined through the parameter $\mu=\vert\min[E_{c}(\mathbf{k})]+\max[E_{v}(\mathbf{k})]\vert/2$
due to the first term in (\ref{hexwarping}). The effective model
$H_{D}+H_{w}+H_{exc}$ can be considered as a good description for
the interaction between the TI surface states and smooth varying ferromagnetic
in-plane magnetization. However, as we argued before there are many
materials where the magnetization varies at the order of the lattice
spacing and the above effective description cannot be directly applied.
Even in such cases, a non vanishing gap can be found if one goes to
the microscopic description of the system. 
\begin{figure}
\includegraphics[scale=1.2]{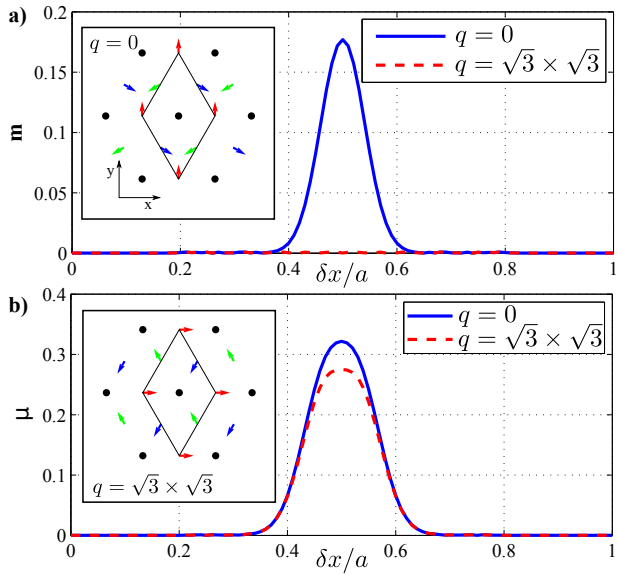} \caption{Evolution of the gap a) and the self doping b) as a function of the
mismatch $\delta\mathbf{x}$ between the two sublattices, as it is
explained in the text. Insets correspond to magnetic configurations
with $\delta\mathbf{x}=0$.}

\label{gapsanddopes} 
\end{figure}

We will exemplify this situation considering the Kagome lattice with
classical planar magnetic configurations. The magnetism on this frustrated
lattice is a current subject of research\cite{R94}. In particular
we have chosen the $q=0$ and $q=\sqrt{3}\times\sqrt{3}$ ground state
spin configurations as plotted in the insets of Fig.(\ref{gapsanddopes}).
In order to monitor the evolution of the spectral properties of the
system we have chosen the lateral displacement $\delta\mathbf{x}$
between lattices in the $OX$ direction as a control parameter. In
this way the magnetic moment sitting on the horizontal axis will play
a dominant role. The results for the gap and for the self doping are
also displayed in Fig.(\ref{gapsanddopes}). We can easily understand
the results by keeping in mind that the exchange interaction between
magnetic moments and electron spins is short ranged and the behavior
of the gap with the magnetic moment components according to the Fu's
model. In the inset of fig.(\ref{gapsanddopes}a) the red sublattice
magnetization points along the $OY$ direction so the gap will open
when this sublattice magnetization is closest to the atomic lattice.
On the contrary, in the inset of Fig.(\ref{gapsanddopes}b) the red
sublattice magnetization points along the $OX$ axis so according
Fu's model, no gap will be generated. Also, it is expected that both
configurations will give rise to a non vanishing self doping effect
when the effective magnetization is non zero as it is shown in Fig.(\ref{gapsanddopes}b).
Another important observation is that different in-plane magnetic
configurations in adjacent space regions (Néel domain walls\cite{M63})
might induce a mass with opposite sign which would generate chiral
1D fermionic states.

\emph{Experimental feasibility.} One of the proposed ferromagnetic
insulators is the EuO\cite{ST02}. It possesses a gap of the order
of $1.2eV$ and it crystallizes in the simple cubic structure, not
commensurate to the triangular lattice structure of the surface, introducing
further complexity in the problem. In contrast to ferromagnetic insulators,
ferrimagnetic insulators offer more reliable experimental opportunities.
The ferrimagnetic insulating state is present
in Nature in many compounds and in many crystalline structures, and
ferrimagnetic thin films can be manufactured in many ways\cite{KP07}.
Among them we highlight the hexagonal ferrites of which PbFe$_{12}$O$_{19}$
is the archetypal material. They crystallize in the hexagonal magnetoplumbite
structure having a rather complex atomic configuration. Many other
ferrites grow in the spinel structure, like the magnetites (Fe$_{3}$O$_{4}$)
and Cobalt ferrites (CoFe$_{2}$O$_{4}$) that might be grown in thin
films with appreciable out-of-plane magnetization\cite{LW07}. Although
CoFe$_{2}$O$_{4}$ have a strong mismatch between the magnetic and
the $Se$ lattice structure and the results presented here are not
directly applicable we suggest it as prospect candidate for experimentally
analyzing the effect of ferrimagnetism on the surface states of a
TI. Concerning in-plane magnetic configurations, we can mention Kagome
systems with different planar spin ground states like SrCr$_{9}$Ga$_{3}$O$_{19}$,
herbersmiththite, jarosite and many others\cite{G01}.

\emph{Conclusions.} In the present paper we have addressed the question
if the effective Hamiltonian (\ref{masslessDirac}) is valid when
the helical surface states of a TI are coupled to magnetically active
layers, . By using a tight-binding model for both the TI and the magnetization,
we have shown that contrary to the (perhaps too) naive expectation
that the helical spin couples to the total magnetization present in
the unit cell, it couples to a \emph{weighted average} of the magnetic
moments present in the unit cell. This result tell us that in principle,
there is no physical reason for ruling out antiferromagnetic insulating
thin films as candidates for inducing gaps in TI surface states. We
have considered also the possibility of ferrimagnetic insulating thin
films. In all the cases, we have shown that the magnetic exchange
mechanism induces a gap in these surface states. As a result, we have
found that the gap is sensitive to the mismatch between the magnetic
and surface lattices. We have considered also the realistic situation
where the film magnetization is in-plane and homogeneous. In this
case, a gap might be opened due to hexagonal warping effects\cite{F09}
even for materials whose thin film magnetization is textured at the
scale of the lattice spacing, inducing a zero average magnetic moment
per unit cell. 
\begin{acknowledgments}
The authors gratefully acknowledge conversations with Carlos Pecharromán,
András Pályi and József Cserti. A. C. aknowledges the CSIC JAE-doc
fellowship program for financial support. O. L. aknowledges the support
of the EU grant NanoCTM. 
\end{acknowledgments}

\bibliography{antiferribiblio}

\end{document}